\pdfoutput=1

\pdfoutput=1

\documentclass[aps,pra,twocolumn,superscriptaddress,longbibliography]{revtex4-2}
\usepackage{graphicx,bm,amsmath,eufrak,amssymb,url}
\usepackage[usenames,dvipsnames]{color}
\usepackage[bbgreekl]{mathbbol}
\usepackage{xcolor}
\usepackage{braket}
\usepackage[colorlinks=true, breaklinks=true, linkcolor=darkblue, citecolor=darkblue, urlcolor=darkblue]{hyperref}
\usepackage{pgf}

\usepackage{subcaption}

\usepackage[T1]{fontenc}
\usepackage[english]{babel} 
\usepackage{verbatim}
\usepackage{graphicx}

\usepackage{bbold}

\newcommand{\trAE}[2][]{\ensuremath{\textnormal{Tr}_{#1}\left[ #2 \right]}}

\definecolor{darkblue}{rgb}{0, 0, 0.8}
\definecolor{bluegreen}{rgb}{0, 179, 184}

\newcommand{\be}{\begin{equation}}
	\newcommand{\ee}{\end{equation}}
\newcommand{\bea}{\begin{eqnarray}}
	\newcommand{\eea}{\end{eqnarray}}

\def\doi{http://dx.doi.org/}

\graphicspath{{./Figures/}}

\begin{document}
	\title{A randomized measurement toolbox for {an interacting Rydberg-atom quantum simulator}}

		\author{Simone Notarnicola}
	\affiliation{Dipartimento di Fisica e Astronomia ``G. Galilei'', Universit\`a di Padova, I-35131 Padova, Italy.}
        \affiliation{Department of Physics, Harvard University, Cambridge, MA 02138, USA}
	\affiliation{Padua Quantum Technologies Research Center, Universit\`a degli Studi di Padova.}
	\affiliation{Istituto Nazionale di Fisica Nucleare (INFN), Sezione di Padova, I-35131 Padova, Italy.}  
	\author{Andreas Elben}
	\affiliation{Institute for Quantum Information and Matter and Walter Burke Institute for
Theoretical Physics, California Institute of Technology, Pasadena, CA 91125, USA}
	\affiliation{Center for Quantum Physics, University of Innsbruck, Innsbruck A-6020, Austria}	
\affiliation{Institute for Quantum Optics and Quantum Information of the Austrian Academy of Sciences,  Innsbruck A-6020, Austria}
	\author{Thierry~Lahaye}
\affiliation{Universit\'e Paris-Saclay, Institut d'Optique Graduate School,
CNRS, Laboratoire Charles Fabry, 91127 Palaiseau Cedex, France}
	\author{Antoine~Browaeys}
\affiliation{Universit\'e Paris-Saclay, Institut d'Optique Graduate School,
CNRS, Laboratoire Charles Fabry, 91127 Palaiseau Cedex, France}
	
	
	\author{Simone~Montangero}
	\affiliation{Dipartimento di Fisica e Astronomia ``G. Galilei'', Universit\`a di Padova, I-35131 Padova, Italy.}   
	\affiliation{Padua Quantum Technologies Research Center, Universit\`a degli Studi di Padova.}
	\affiliation{Istituto Nazionale di Fisica Nucleare (INFN), Sezione di Padova, I-35131 Padova, Italy.}   
	
		\author{Beno\^it Vermersch}
	\affiliation{Universit\'e  Grenoble Alpes, CNRS, LPMMC, 38000 Grenoble, France}
\affiliation{Center for Quantum Physics, University of Innsbruck, Innsbruck A-6020, Austria}	
\affiliation{Institute for Quantum Optics and Quantum Information of the Austrian Academy of Sciences,  Innsbruck A-6020, Austria}

	\date{\today}
	
	\begin{abstract}
We present a toolbox to probe quantum many-body states implemented on Rydberg-atoms quantum hardware
via randomized measurements.
We illustrate the efficacy of this measurement toolbox in the context of probing entanglement, via the estimation of the purity, and of verifying a ground-state preparation using measurements of the Hamiltonian variance.
To achieve this goal, we develop and discuss in detail a protocol to realize independent, local unitary rotations. We benchmark the protocol by investigating the ground state of the one-dimensional SSH model, recently realized on a chain of Rydberg atom, and the state resulting after a sudden quench in a staggered XY chain. We probe the robustness of our toolbox by taking into account experimental imperfections, such as pulse fluctuations and measurement errors.
	\end{abstract}
	
	\maketitle
	
	\section{Introduction}\label{sec:introduction}
	Synthetic quantum systems, composed of, e.g., neutral atoms~\cite{Gross2017,Browaeys2020}, ions~\cite{monroe2021programmable}, superconducting qubits~\cite{Kjaergaard2020}, allow us to engineer spin-lattice models or implement quantum algorithms on qubit registers, 
	with precise control over geometry and  interactions.
    Among these platforms,  Rydberg atoms have emerged as a promising system. 
	They can be described in good approximation in terms of qubits, with the spin-up state $\ket{\uparrow}\equiv \ket{1}$ encoded by a Rydberg state, and the spin-down $\ket{\downarrow}\equiv \ket{0}$ encoded by another Rydberg state, or an atomic ground state~\cite{Browaeys2020}.  One of the most  relevant assets for Rydberg quantum technologies is the long qubit lifetime, which scales as $n^3$, where $n\sim 50-100$ is the atom principal quantum number. 
    In addition,  interactions between Rydberg qubits
	are naturally obtained via the dipole-dipole interactions, whose characteristic energy scales as $n^4$ in the resonant regime, and $n^{11}$ in the off-resonant van der Waals regime. 
	Finally, Rydberg atoms can be placed on almost arbitrary geometries using optical forces generated, e.g., by optical tweezers~\cite{Barredo2018}.
	In particular, recent experimental progresses in this direction allowed to experimentally study strongly correlated quantum states with hundreds of qubits in two dimensional lattice models
    ~\cite{scholl2021quantum,ebadi2021quantum}.
    On the quantum computing side, Rydberg atom platforms have demonstrated remarkable performances in terms of scalability, connectivity, and gate fidelities~\cite{levine2019parallel,Madjarov2020high,Bluvstein2022}.
	
				\begin{figure}  
		\centering
		\includegraphics[width=0.9\columnwidth]{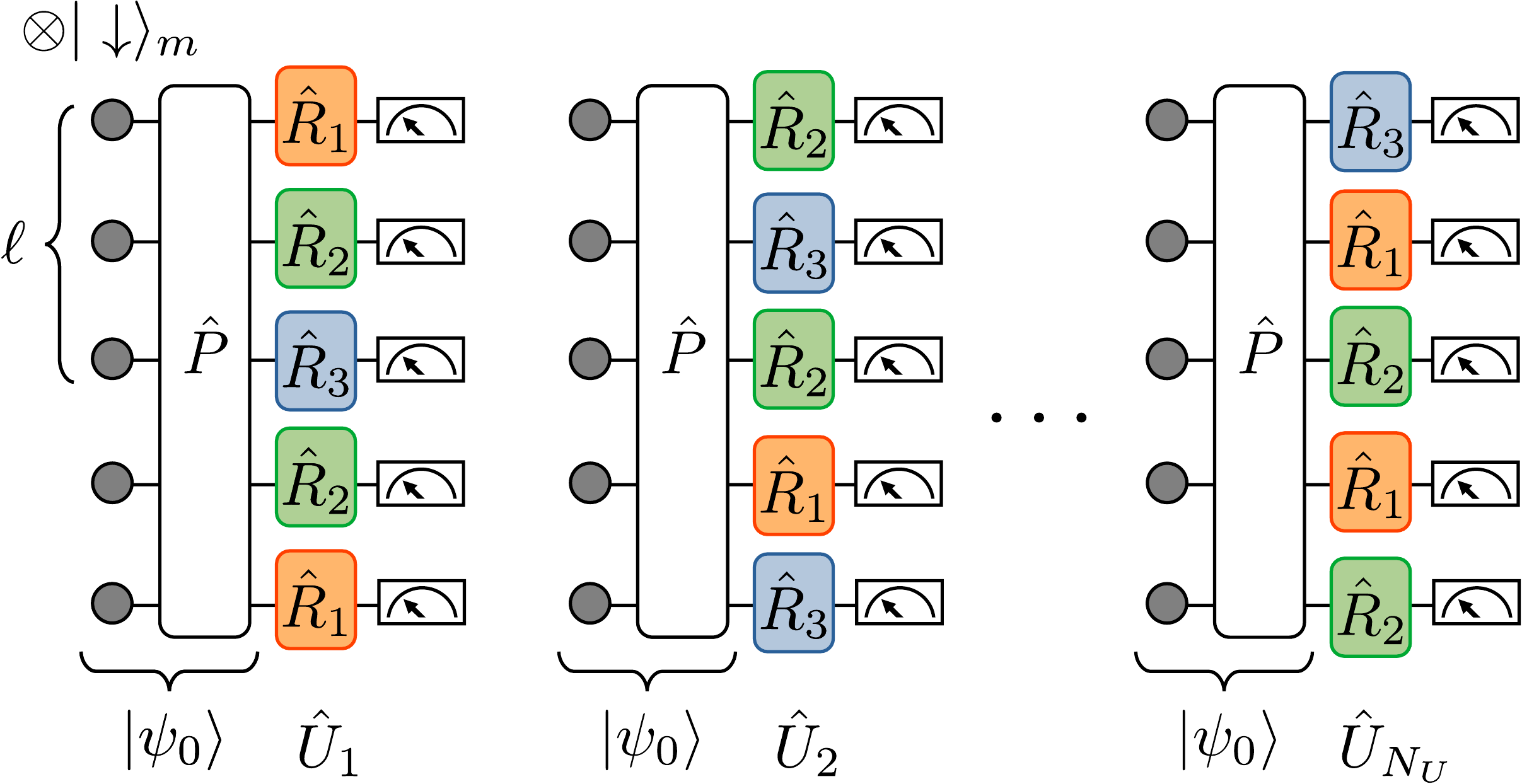}
		\captionsetup{justification=centerlast}
		\caption{ \label{fig:prot_intro}
			Pictorial representation of the randomized measurements toolbox. Starting from a separable state (all sites in $\ket{\downarrow}$), an adiabatic state preparation ${P}$ is applied to implement the target state of interest $|\psi_0\rangle$~\cite{DeLeseleuc2019}. Then, we randomly choose and apply loal unitaries for each lattice site, repeating the protocol for $N_U$ samplings. For each sampling ${U}_i$, we estimate the probabilities $P_{U_i}(\mathbf{s})$ (see Eq.~\eqref{eq:rhoO}) estimating the Hamiltonian variance of $|\psi_0\rangle$ and the purity of system bipartitions with size $\ell$. 
		}
	\end{figure} 
	
	In order to take advantage of all the promising aspects of Rydberg atoms for quantum technologies, it is desirable to equip such platforms with a \emph{measurement toolbox} to extract key physical quantities, such as fidelities and entanglement, and in a state-agnostic way.
	A promising approach in this context consists in using randomized measurements (RMs) based on performing random single-qubit rotations followed by measurements in the computational basis~\cite{VanEnk2012,Elben2018,Huang2020} (see Figure~\ref{fig:prot_intro}).
	
	Randomized measurements (RMs)~\cite{Elben2023}	have  been used to estimate the purity and the second R\'{e}nyi entropy $S_2=-
	\log \mathrm{Tr}[\rho^2] $ of (sub-)systems consisting of up to ten qubits in a trapped ion~\cite{Brydges2019,Brydges2019,Joshi2020,Elben2020_Mixed,Zhu2022} and superconducting qubit quantum simulator~\cite{Vovrosh2021,Satzinger2021}.
	An alternative purity estimator can be obtained via the classical shadows formalism~\cite{Huang2020,Elben2020_Mixed} (see also below). In addition, it has been proposed to use randomized measurements to reveal other properties  of many-body quantum states. This concerns for instance the fidelity of quantum states realized in different experiments~\cite{Elben2020_Mixed,Zhu2021} and versus an ideal theoretical target state, as well as many-body topological invariants associated with (symmetry-protected) topological phases~\cite{Elben2020_SPT,Cian2020}. Moreover, beyond the immediate opportunity to extract via RMs entanglement entropies and related quantities for quantum simulation, our toolbox enables the measurement of arbitrary observables based on the classical shadow formalism~\cite{Huang2020}.
	This is in particular relevant
	in the context of variational quantum optimization algorithms~\cite{farhi2014quantum,peruzzo2014},
	where an observable cost function is repetitively measured.
	

{In the aforementioned ion implementation of the local random unitaries, interactions were absent during the local rotations protocol. Analogously, recent Rydberg atom quantum simulation and computation platforms exploit a qubit implementation in which the $|0\rangle$ and $|1\rangle$ states correspond to hyperfine ground states~\cite{evered2023highfidelity, Bluvstein2022, PhysRevResearch.4.033019, jaschke2022abinitio, Jandura2022timeoptimaltwothree}. In this case, it is possible to locally manipulate the quantum many-body state in the absence of atom-atom interactions.} 

{In this work, we answer the question of whether the local unitaries apparatus can also be exploited in cases when the atom-atom interactions cannot be avoided. To do so, we equip an interacting Rydberg atom quantum simulator with a tailored RM toolbox based on local unitaries, i.e., spin qubit rotations. Despite the protocol we present constitutes a receipt tailored for a particular Rydberg atom setup, it nevertheless demonstrates that the local unitaries apparatus can also be extended to the case of qubits where interactions cannot be turned off}. 
	Our approach complements a recent work employing quasi-local `scrambled' unitaries generated by Rydberg interactions and used for benchmarking and fidelity estimation \cite{Choi2023}. 
    Our approach fully relies on experimental tools that are currently available.
	It consists in implementing RMs via random single-qubit rotations combining local light shifts and time-varying global microwave drives to the atoms. 
 Our approach proposes directly operating on the Rydberg manifold, allowing to realize random single qubit rotations in parallel, i.e.,  with a duration that does not scale with the number of qubits. One can also straightforwardly adapt our protocols for hyperfine state qubits, e.g., replacing microwave drives by Raman pulses.
 
Importantly, we show that the effects of the interactions between Rydberg atoms during the generation of such single-qubit rotations can be made negligible using an optimized pulse sequence for the different drives.
We illustrate the use of the RM toolbox for the measurement of the purity $p_2$  [giving access to the entanglement R\'enyi entropy $S_2=-\log_2(p_2)$] in the SSH model, and of the Hamiltonian variance, which can be used to verify experimentally ground state preparation~\cite{Kokail2020,Huang2020}. {We also estimate the purity of a state resulting from the dynamics after a sudden quench with a staggered XY model.}
Our simulations take into account the most important realistic error sources and analyze the role of statistical errors. We conclude that RMs can be implemented via the presented approach in existing Rydberg platforms.

	In the following, we describe our general RM toolbox in Sec.~\ref{sec:protocol} and propose an experimental implementation in Sec.~\ref{sec:exp}.
	In Sec.~\ref{sec:ssh}, we illustrate our approach in the context of characterization of topological phases with entanglement entropies. We also show the measurement of the Hamiltonian variance of the SSH model, in order to verify the adiabatic preparation of the ground state. 
	
	\section{Local random unitaries toolbox}\label{sec:protocol}
	
	
	\subsection{Probability estimation}
	
	Randomized measurements provide a powerful toolbox to investigate the properties of quantum many-body systems beyond standard low-order correlation functions \cite{VanEnk2012,Elben2018,Vermersch2018,Brydges2019,Vermersch2019,Elben2019,Ketterer2019,Elben2020_xPlatform,Elben2020_SPT,Huang2020,Elben2020_Mixed,Cian2020,Zhou2020,10.21468/SciPostPhys.12.3.106,Garcia2021,PhysRevLett.127.260501,Yu2021,Neven2021,Knips2020,Imai_2021,Rath2021a,PhysRevX.12.011018}. In the following, we outline the randomized measurement protocol employing local (single-spin) random unitary operations. To this end,  we consider a quantum state $\rho$ defined on a lattice of qubits with size $L$ and associated Hilbert space of dimension $2^L$. We denote its computational $\mathbf{z}$-basis with $\{\ket{\mathbf{s}}\}$  with bitstrings $\mathbf{s}=(s_1,\dots,s_L)$ and $s_m =0,1$ for $m=1,\dots, L$. 
	A randomized measurement comprises the following steps: (i) A random unitary $ U=\bigotimes_{m=1}^L u_m$ is applied to $\rho$, where each $u_m$ is sampled independently from an appropriate ensemble of local (single-spin) unitary transformations, typically a  unitary $2$-design \cite{Dankert2009,Gross2007}.  Examples of such unitary 2-designs include continuous single-spin rotations which cover the Bloch sphere of each spin uniformly [the Haar measure on the unitary group $U(2)$] as well as  the (discrete) single-qubit Clifford group \cite{Dankert2009,Gross2007}. (ii) This is followed by a measurement in the computational $\mathbf{z}$-basis with outcome bit-string $\mathbf{s}=(s_1,\dots,s_L)$. This sequence is then rerst with the same unitary $U$ to obtain an estimate of the probabilities $P_U(\mathbf{s})=\trAE{U \rho U^\dagger \ket{\mathbf{s}}\bra{\mathbf{s}}}$, and subsequently with newly sampled unitaries to estimate the average over the ensemble of unitary transformations. 
	We denote the number of repetitions with the same  random unitary with $N_{meas}$  and the number of applied unitaries $U$ with $N_U$ such that the rotations protocol is repeated $N_{tot}=N_U \times N_{meas}$ times in total.

	In this work, we choose to use local random unitary operations $u_m$  which are sampled from the discrete, finite single-qubit Clifford group $U(2)$. Since for randomized measurements, the application of a random unitary is directly followed by a computational $\mathbf{z}-$basis measurement in the $\mathbf{z}$-direction, the application of randomly sampled single-qubit Clifford gates is equivalent to sampling the $L$ measurement directions $\mathbf{v}_m$ among a finite set of three mutually orthogonal directions, as for example $\{\mathbf{x},\mathbf{y},\mathbf{z}\}$ (see 
	Ref.~\cite{Huang2020}).
	In this case, the corresponding set of transformations is $\mathbf{R}=\{\mathrm{e}^{-i \pi/4 \sigma_y},\mathrm{e}^{-i \pi/4 \sigma_x},\mathbf{1}\}$, that rotates each direction onto the measurement axis $\mathbf{z}$.

	\subsection{Purity estimation}
	
	A key application of randomized measurements is the estimation of the purity of quantum states to characterize the coherence of the underlying quantum device and to reveal entanglement \cite{VanEnk2012,Elben2018,Vermersch2018, Brydges2019,Elben2019,Huang2020,Rath2021}. In the following, 
	we consider a system with size $L$ and with basis $\{\mathbf{s}\}$, and a sub-system $A$ with size $N_A$. 
	The purity $\trAE[]{\rho_A^2}$ of the reduced density matrix $\rho_A$ of $A$ can be estimated, following the procedure presented in Refs.~\cite{Brydges2019,Elben2019}. Given the estimates of the probabilities $P_U(\mathbf{s})$, one obtains estimates of the probabilities  $P_U(\mathbf{s}_A)=\sum_{\mathbf{s}|_A=\mathbf{s}_A}P_U(\mathbf{s})$ of computational basis states $\ket{s_A}$ for  any subsystem $A$ via post-processing. Then, the purity $\trAE[]{\rho_A^2}$ is obtained from second-order correlations of the probabilities  $P_U(\mathbf{s}_A)$ via
	\begin{equation}\label{eq:Tr2}
		\mathrm{Tr}[\rho_A^2] = 2^{N_A} \sum_{\mathbf{s}_A,\mathbf{s}_A'}(-2)^{-D[\mathbf{s}_A,\mathbf{s}_A']}\overline{P_U(\mathbf{s}_A)P_U(\mathbf{s}_A')} \,.
	\end{equation}
	Here, $D[\mathbf{s}_A,\mathbf{s}_A']$ denotes the Hamming distance of the bitstrings $\mathbf{s}_A$ and $\mathbf{s}_A'$ and $\overline{\vphantom{h}\dots}$ the ensemble average over the local random unitaries. 
	  Eq.~\eqref{eq:Tr2} represents an exact relation in the limit of $N_{meas}\rightarrow \infty$ and when the local random unitaries are averaged over a complete unitary $2$-design. In practice, statistical errors arise from a finite number of measurements $N_{meas}$ per unitary and a finite number $N_U$ of local random unitaries sampling the ensemble average. Numerical and analytical analysis of such  statistical errors showed that the total number of experimental runs $N_U N_{meas}$ to estimate the purity with high confidence {and probability} scales approximately  as $2^{b N_A}$ with $b\approx 1$.  The exact value of $b$ and the optimal ratio $N_U/N_{meas}$  depends on the state of interest and the required precision. {We note that this represents a substantial improvement compared to full quantum state tomography, requiring at least $2^{b' N_A}$ experimental runs with $b'\gtrsim 2$ (see e.g.\ Ref.~\cite{guta2020}).} The scaling of statistical errors with system size can furthermore be substantially improved  via importance sampling~\cite{Rath2021}.

	\subsection{Estimating expectation values of arbitrary observables}
	
	The  same randomized measurement data
	can be used to estimate expectation values $\trAE{O\rho}$ of arbitrary observables $O$  \cite{Huang2020}. 
	Utilizing the tomographic completeness of randomized measurements \cite{Ohliger_2013, Elben2019, Huang2020}, the expectation value of arbitrary observables $O$ can be obtained via
	\begin{align}
		\trAE[]{\rho O} = 2^N \sum_{\mathbf{s},\mathbf{s}'} (-2)^{-D[\mathbf{s},\mathbf{s}']} \overline{P_U(\mathbf{s}) \braket{\mathbf{s}'|UOU^\dagger|\mathbf{s}'}} .\label{eq:rhoO}
	\end{align}
	Differently from Eq.~\eqref{eq:Tr2}, this expression is linear in the experimentally estimated outcome probabilities $P_U(\mathbf{s})$. Hence, the procedure to estimate expectation values  $\trAE[]{\rho O}$ is as follows: In the experiment, we estimate outcome probabilities $P_U(\mathbf{s})= \braket{\mathbf{s}|U\rho U^\dagger|\mathbf{s}}$, as in the case of the purity estimation. On a classical computer, we calculate the corresponding matrix elements  $\braket{\mathbf{s}|UOU^\dagger|\mathbf{s}}$ (for the same unitaries $U$ which have been applied in the experiment).  Then, we cross-correlate according to Eq.~\eqref{eq:rhoO}.
	
	Observable estimation with randomized measurements has been formalized and rigorous error bounds have been obtained via the classical shadows formalism \cite{Huang2020}. There, it has been shown that statistical errors depend on the set of observables $O$ of interest. Below, we consider the specific case of $O=H$ ($O=H^2$)  with $H$ being a  Hamiltonian with 
	$k$-body interactions.  Then, in the limiting case $N_{meas}=1$, $N_U \sim 2^k \log(N)$  ($N_U \sim 2^{2k} \log(N^2)$)  random unitaries  are required to estimate $\trAE{H  \rho}$ ($\trAE{H^2\rho}$) with high confidence and probability \cite{Huang2020}. 
	This number can be substantially further decreased using derandomization techniques~\cite{Huang2021}. We furthermore note that the Eq.~\eqref{eq:rhoO} can be generalized to estimate expectation values of arbitrary multi-copy observables~\cite{Huang2020}. This enables, for instance, the detection of  mixed state entanglement via higher-order moments of (the partial transpose) of the density matrix $\rho$~\cite{Elben2020_Mixed}
	(see also Ref.~\cite{Zhou2020}), of symmetry-resolved entanglement entropies~\cite{Vitale2021}, and of the quantum Fisher information~\cite{Rath2021a}.

	In contrast to the purity estimation formula, in Eq.~\eqref{eq:rhoO} we utilize  explicitly the knowledge of the random unitaries $U$ to calculate the required matrix elements $\braket{\mathbf{s}|UOU^\dagger|\mathbf{s}}$. 
	Thus,  
	any miscalibration between the local random unitaries actually applied in the experiment and those 
	applied on the classical computer affects the estimation of $\trAE{\rho O} $ \cite{Elben2020_xPlatform,Chen2021}. We will discuss the influence of such implementation errors in detail below. 
	In addition, the robustness can be improved via calibration experiments with simple  states  which can be prepared with high fidelity~\cite{Chen2021,Koh2022classicalshadows,PhysRevLett.127.030503}. 
	In the next section, we describe our Rydberg quantum optics model, and the corresponding implementation of randomized measurements.

	\section{Proposal to experimentally  implement  the Randomized measurements  toolbox}\label{sec:exp}
\subsection{The model}	
We consider an array of atoms (either one-dimensional as shown here, or two-dimensional), made to interact by exciting them to Rydberg states~\cite{Browaeys2020}.  
	In particular, we focus on the setup used to observe symmetry-protected topological phases in a Su-Schrieffer-Heeger (SSH) chain (see, e.g, Ref.~\cite{DeLeseleuc2019}).
	 By encoding pseudo-spin-$1/2$ states in two dipole-coupled Rydberg levels (such as $nS$ for $\ket{\downarrow}\equiv \ket{0}$ and $nP$ for $\ket{\uparrow}\equiv \ket{1}$, with $n\sim 60$), the dipole-dipole interaction at work between the atoms implements the XY spin Hamiltonian $\sum_{i<j}J_{ij}\sigma^+_i\sigma^-_j+{\rm h.c.}$, with $J_{ij}$ decaying as $1/r_{ij}^3$ with the distance $r_{ij}$ between the atoms $i$ and $j$, and $\sigma^\pm=(\sigma_x\pm i \sigma_y)/2$ are linear combinations of the usual Pauli matrices.  
	
\begin{figure}
		\begin{center}
			\includegraphics[width=85mm]{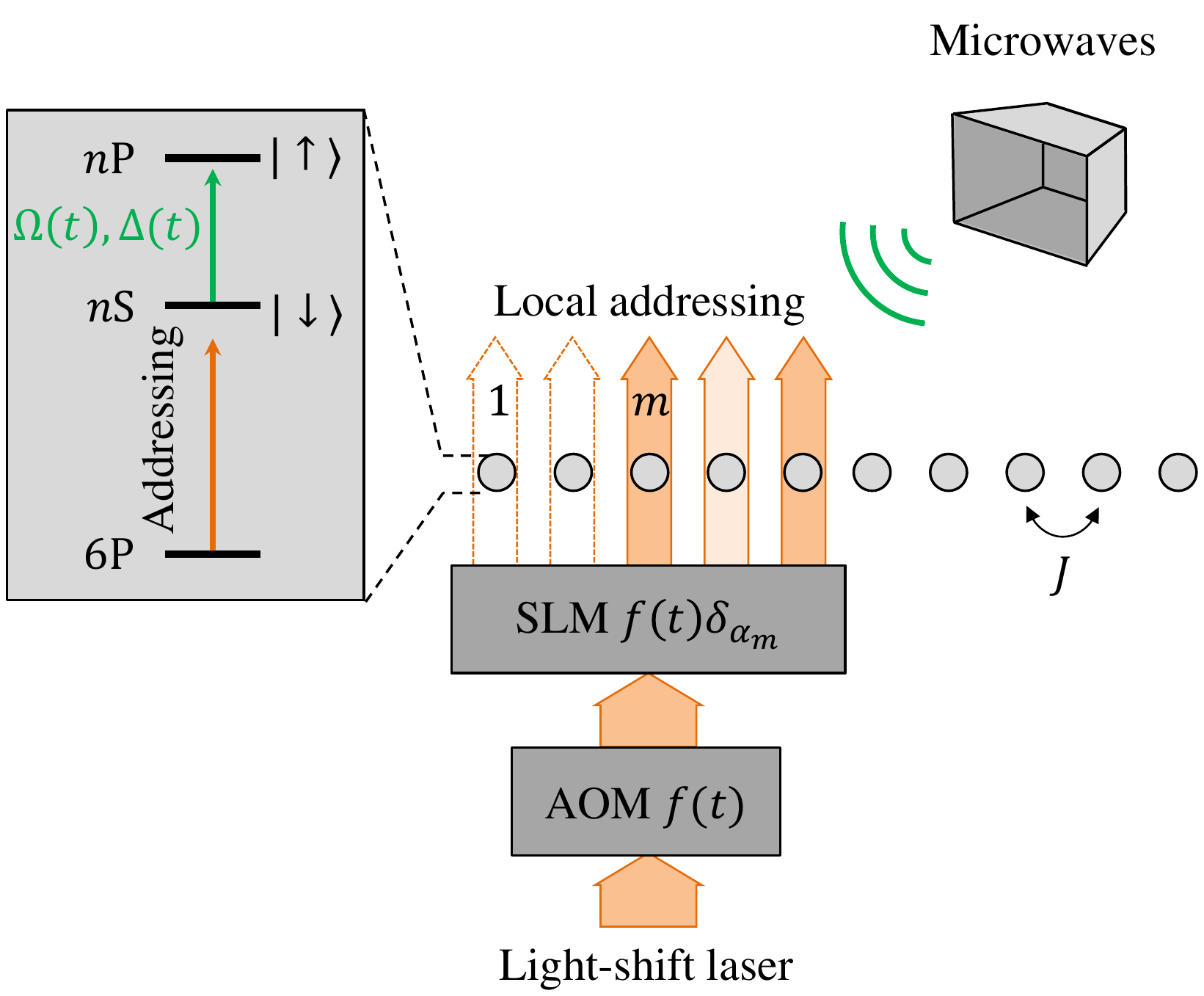}
			\captionsetup{justification=centerlast}
			\caption{The experimental scheme under consideration. The atoms in the tweezers array (here in one dimension) are used to encode a pseudo-spin $1/2$ in two Rydberg states $nS$ and $nP$. Global manipulation of the spin states is achieved using microwave driving with Rabi frequency $\Omega(t)$ and detuning $\Delta(t)$. Additional, site-dependent detunings $\delta_{\alpha_m}$ with a global time dependence $f(t)$ are obtained by using a light-shifting laser beam, controlled by an acousto-optic modulator (AOM) and a spatial light modulator (SLM), to couple off-resonantly the $nS$ state to a low-lying state (such as the $6P$ state for Rb atoms). 
			}
			\label{fig:setup}
		\end{center}
	\end{figure}

	To manipulate the internal spin states and thus implement local rotations, we start from the experimental setup used in Ref.~\cite{DeLeseleuc2019}, sketched in Figure \ref{fig:setup}. In particular, we can first manipulate them \emph{globally} by using microwave pulses with a Rabi frequency $\Omega(t)$ and a detuning $\Delta(t)$. 
	For \emph{local} manipulation, we add a local light shift with a tightly focused laser beam (for instance coupling off-resonantly the $nS$ state to a low-lying $P$ state such as the $6P$ state for Rb) on a selected atom in order to tune the qubit frequency into (or out of) resonance with the microwave field \cite{DeLeseleuc2017}. A spatial light modulator (SLM) is used to program at will the spatial dependence of these addressing beams, while the (global) time dependence $f(t)$ of the intensity of the addressing beams is set with an acousto-optic modulator (AOM), placed before the SLM, and that allows for the generation of fast pulses.

	By taking all the available terms into account, the experimental Rydberg Hamiltonian describing the local transformations is 
	\begin{equation}\label{eq:prot_ham}
		{H}_{prot}(t)=\sum_{m=1}^L
		\left[\frac{\Omega(t)}{2}\sigma_m^x - [ \Delta(t)-f(t)\delta_{\alpha_m}] {n}_m\,
		\right],
	\end{equation}
	where $\sigma^\alpha_m$ are the Pauli matrices acting on site $m\in [1,L]$, $\alpha_m\in\{1,2,3\}$ and ${n}_m=(\sigma_m^z+\mathbf{1}_m)/2$. 

	During the application of the rotation protocol, the state is evolving under the total Hamiltonian
	\begin{equation}\label{eq:tot_ham}
		{H}(t)={H}_{prot}(t)+{H}_{mod}\,,
	\end{equation}
	where ${H}_{mod}$ is the static model Hamiltonian describing the interactions between Rydberg atoms, see Eq.~\eqref{eq:ham_mod} below.
	  The interaction terms can create spurious correlations between the local rotations and affect the estimation of the probabilities $P_U(\mathbf{s})$. In the following, we determine the pulses of the rotation protocol after fixing the maximum amplitude of the Hamiltonian ${H}_{mod}$ parameters $J$ to be $< 1\,\mathrm{MHz}$.

\subsection{Experimental proposal}

In the following, we exploit the Hamiltonian $ H_{prot}(t)$ to simultaneously implement three transformations
${\mathbf{R}}=\{{R}_1,{R}_2,{R}_3\}$ that rotate the measurement axis $\mathbf{z}$ onto three mutually perpendicular directions. 
    First, we consider the idealized limit of very short {protocol time} $T$ (i.e. { $J T\ll 1$, to} neglect the influence of interactions)  and {an} arbitrarily  large detuning $\Delta(t)$ much larger than the Rabi frequency $\Omega(t)$. The local potentials $f(t) \delta_{\alpha_m}\sim \Delta(t)$ are fixed to implement the different rotations.
In this limit, we can present  analytical pulse sequences to realize {$\mathbf{R}=\{\mathrm{e}^{-i \pi/4 \sigma_x},\mathrm{e}^{-i \pi/4 \sigma_y},\mathbf{1}\}$} exactly.
{We consider constant, square pulses. The Rabi frequency pulse $\Omega(t)$ has a duration of $T$. It implements two successive $\pi/2$ rotations and satisfies the condition $\Omega(t)T/2=\pi/2$. Instead, the detuning term $\Delta(t)$ is null in the first half of the protocol, while we set $\Delta(t) T/2=\pi/2$ in the second half (this choice will be clarified in the following).}

The transformation ${R}_{1}$ is implemented by setting $T f(t) \delta_1=0$. {During the first half of the total time interval $T$, we have $T \Delta(t)=0$ and the Rabi frequency term implements a $\pi/2$ pulse.} %
In the second half-time, the detuning pulse by $\Delta(t)\gg \Omega(t)$ shifts the transition between the two levels off-resonant. Note that the $\Delta(t)$ pulse adds a phase that does not affect the measurement along the $\mathbf{z}$ axis. 
	In order to implement the ${R}_{2}$ transformation, we decompose the rotation around the $\mathbf{y}$ axis into the sequence of rotations $\mathrm{e}^{-i \pi/4 \sigma_z}\mathrm{e}^{-i \pi/4 \sigma_x}\mathrm{e}^{i \pi/4 \sigma_z}$. Then, we ignore the last $\sigma_z$ rotation which does not influence the final measurement outcome, and resulting in the rotations {$\mathrm{e}^{-i \pi/4 \sigma_x}\mathrm{e}^{i \pi/4 \sigma_z}$} to be implemented. 
	The rotation around $\mathbf{z}$ is realized by setting $T f(t)\delta_2/2 = \pi/2$, with $f(t)\delta_2 \gg \Omega(t)$, in the first half of the protocol. {In the second half, the choice of $\Delta(t)$ realizes the condition $\Delta(t)-f(t)\delta_2 = 0$}, and the pulse $\Omega(t)$ implements the rotation around $\mathbf{x}$.
	Finally, the rotation ${R}_{3}$ is realized by the pulse with amplitude $f(t) \delta_3\gg \Omega(t)$ in the first half of the protocol and $\Delta(t) - f(t)\delta_3 \gg \Omega(t)$ in the second half (recall that $T \Delta(t)=0$ during the first half of the protocol).  {The effects of $\Omega(t)$ can be neglected as the pulse $f(t)\delta_3$ always guarantees the off-resonance condition.}

	Starting from this ideal setting described above, we now assess the role of finite time preparation and amplitude parameters. We keep the same general pulse sequence and investigate whether it implements the required transformations $\mathbf{R}$ with high fidelity. In particular, the pulses implemented by the functions $\Omega(t)$ and $\Delta(t)$ are the same in all three cases, while the amplitudes $\delta_\alpha$ can be different and are set to three values $\delta_1,\,\delta_2,\,\delta_3$.
	The figure of merit we use to search the pulses is
\begin{equation}\label{eq:fidelity}
	\mathcal{A}_{\alpha}(\mathbf{R})=|\varepsilon_{\alpha\beta\gamma}\langle\uparrow|{R}_{\beta}{R}_{\gamma}^{^\dag}|\uparrow\rangle|^2=1\;  \,\alpha\in\{1,2,3\},
\end{equation}
where $\varepsilon$ is the antisymmetric Levi-Civita Tensor, and indexes $\beta,\, \gamma$ are implicitly summed. The $\sigma_z$ eigenstate $\ket{\uparrow}$ is one of the two possible measurement outcomes.
Note that for short pulses, larger than the interactions strength, we compute the figure of merit on one-site rotations, independently on the size of the system. We numerically investigate the role of interactions below.
To test each parameters choice,	 we consider $N_{tot}=10^5$ copies of the local protocols and add random fluctuations in each pulses realization. We model them as independent Gaussian fluctuations. We assume their variance to be proportional to the pulse amplitude through the percentage coefficient $\varepsilon_\%=3$ (the effect of considering different coefficients will be analysed in the following).
	As a result, we obtained the protocol $\mathbf{R}^*$ shown in Figure \ref{fig:totprot}, for which  $\mathcal{A}_k(\mathbf{R}^*)/2=\{0.55\pm0.06,\,0.56\pm0.05,\,0.58\pm0.04\}$.
 
{On the one hand, we chose the shape of the pulses to optimize the figure of merit $\mathcal{A}_{\alpha}(\mathbf{R})$ defined in Eq.~\ref{eq:fidelity}. On the other, we determined their duration and amplitude to minimize the effects of interactions. By fixing the total phase of the pulses, we balanced between the minimum duration of the pulse and the maximum pulse increasing speed. We got a rotation protocol time $T_R \simeq 0.15 \mu \mathrm{s}$, with the largest pulse amplitude of $\sim 7 \mathrm{MHz}$ and an increasing speed of $\sim 35\mathrm{Mhz}/10\mathrm{ns}$. The condition $T_R\ll 1/J_e \simeq 2\mu\mathrm{s}$, with $J_e$ being the largest coupling in the Hamiltonian defined in subsection~\ref{subsec:model}, allows to control the interactions spurious effects}.

To benchmark the rotations $\mathbf{R}^*$, we consider the SSH chain with size $L$. Each random unitary ${U}$ is sampled by randomly attributing to each atom $m$ a label $\alpha_m=1,2,3$  with equal probability $1/3$, corresponding to a parametrization of the system with light shift $f(t)\delta_{\alpha_m}$.
	 	  	 	  To make our analysis realistic, we include measurement errors. In particular, we assume the probability of errors occurring during the readout process. 
	We  model them as a $1\%$ error to detect a false $|\uparrow\rangle$ state and $3\%$ error to detect a false $|\downarrow\rangle$ state~\cite{PhysRevA.97.053803}. 
	We set $N_{tot}=N_U\times N_{meas}<10^5$ to make our estimations compatible with experimental typical capabilities.
	
	\section{Numerical illustration with the SSH model}\label{sec:ssh}
	
	\subsection{Presentation of the model and measured quantities}\label{subsec:model}
	As a testbed, we consider the SSH 1D chain described by the Hamiltonian
	\begin{align} \label{eq:ham_mod} 
		& H_{mod}=\, \\ \nonumber
		&-J_e\sum_{even \,x}\sigma^+_x\sigma^-_{x+1}
		-J_o\sum_{odd \,x}\sigma^+_x\sigma^-_{x+1}+\mathrm{H.c.}+{H}_{nnn}
	\end{align}
	with $(J_e,J_o)=(0.484, -0.18)\, \mathrm{MHz}$. By arranging the atoms as in \cite{DeLeseleuc2019}, next-nearest neighbor interactions are suppressed, while ${H}_{nnn}=-J_{nnn}\sum_{x}\sigma^+_x\sigma^-_{x+3}+\mathrm{H.c.}$ describes spurious next-next nearest neighbor exchange terms, with $J_{nnn}\simeq 0.04\, \mathrm{MHz}$.  We neglect residual van der Waals interactions between excited atoms, while we artificially break the degeneracy of the ground state by adding a local chemical potential term to one of the extreme sites of the chain.
	Starting from the experimental realization of this model Hamiltonian, we apply the toolbox, and show that we can access particular entanglement entropies and Hamiltonian variances.
	\begin{figure}
		\includegraphics[width=0.49\columnwidth]{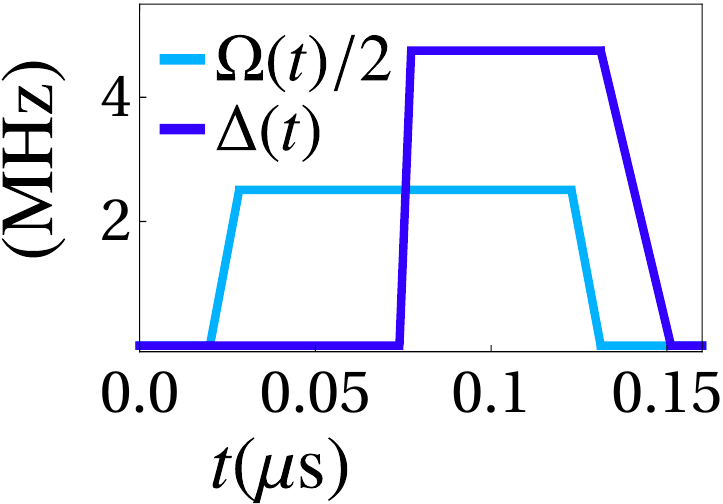} 
		\hfill
	\includegraphics[width=0.49\columnwidth]{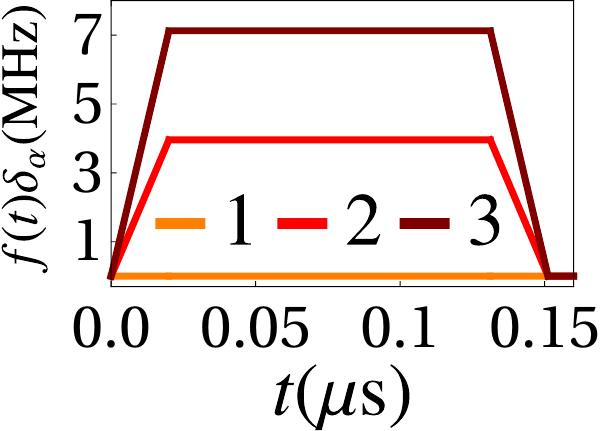}
	\pgfputat{\pgfxy(-4.1,0.4)}{\pgfbox[left,bottom]{(a)}}
	\pgfputat{\pgfxy(0.2,0.4)}{\pgfbox[left,bottom]{(b)}}
		\captionsetup{justification=centerlast}
		\caption{
			\label{fig:totprot}
			Pulses relative to the uniform (a) detuning $\Delta(t)$ and Rabi coupling $\Omega(t)$ and to the space dependent detunings $f(t)\delta_\alpha$ (b).
			The amplitudes $\delta_\alpha$  change to realize the three different rotations $\mathbf{R}^*$.
		}
	\end{figure}

\begin{figure}
	\begin{subfigure}{\columnwidth}
			\captionsetup{position=top,singlelinecheck=off,justification=raggedright}
			\caption{}
			{\vskip -6mm}
			\centering
		\includegraphics[width=0.77\textwidth]{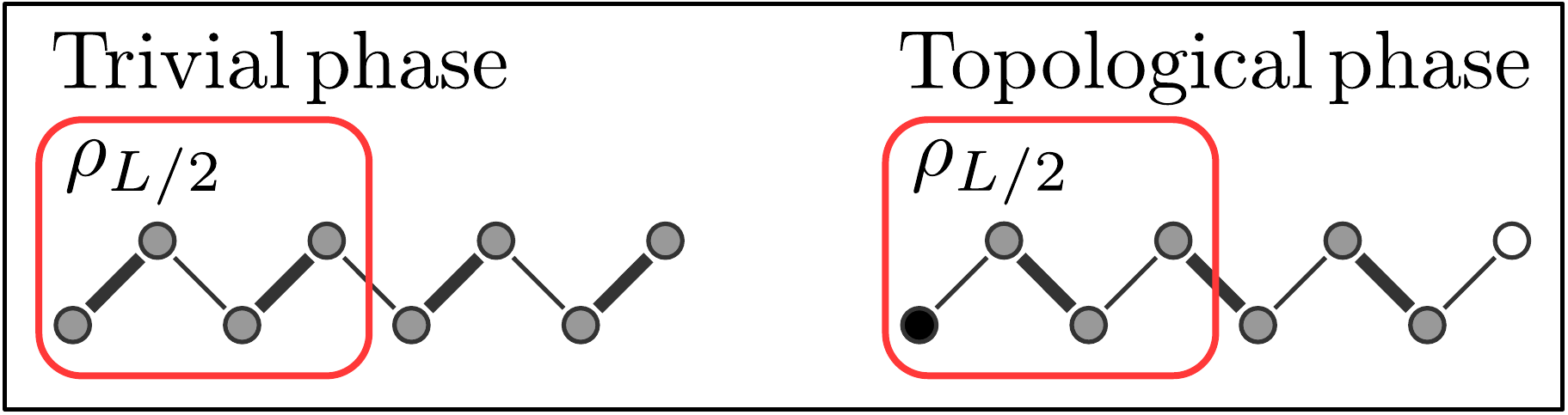}
			\label{fig:sshgs}
		\end{subfigure}
		{\vskip 1.5mm}
		\begin{subfigure}{0.49\columnwidth}
			\captionsetup{position=top,singlelinecheck=off,justification=raggedright}
			\caption{}
			\centering
			{\vskip -4mm}
			\hspace{-3mm}
			\includegraphics[width=\columnwidth]{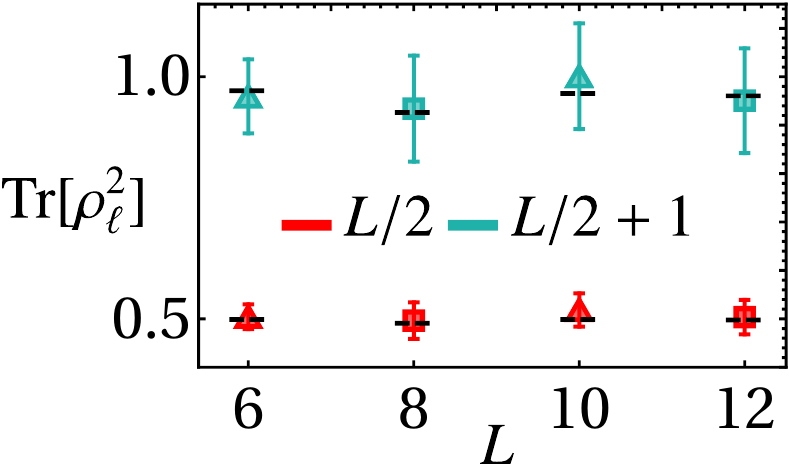}
			\label{fig:tr2_vs_L_teor}
		\end{subfigure} 
		\begin{subfigure}{0.49\columnwidth}
			\captionsetup{position=top,singlelinecheck=off,justification=raggedright}
			\caption{}
			\centering
			{\vskip -4mm}
			\hspace{-3mm}
			\includegraphics[width=\columnwidth]{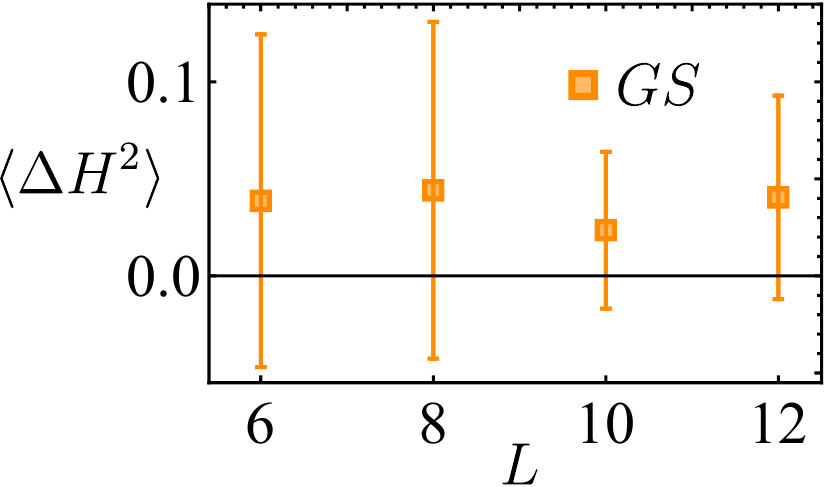}
			\label{fig:H2_vs_L_teor}
		\end{subfigure} 
		{\vskip 0.5mm}
			\begin{subfigure}{0.49\columnwidth}
			\captionsetup{position=top,singlelinecheck=off,justification=raggedright}
			\caption{}
			\centering
			{\vskip -4mm}
			\hspace{-3mm}
			\includegraphics[width=\textwidth]{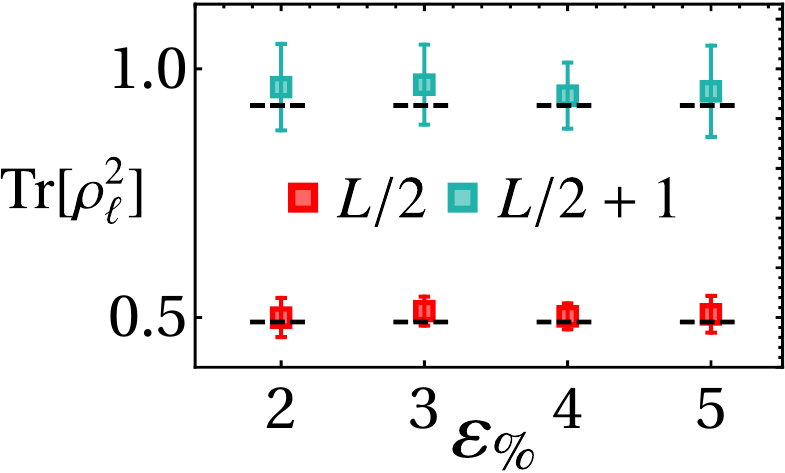}
			\label{fig:tr2_vs_eps_teor}
		\end{subfigure} 
			\begin{subfigure}{0.49\columnwidth}
			\captionsetup{position=top,singlelinecheck=off,justification=raggedright}
			\caption{}
			\centering
			{\vskip -4mm}
			\hspace{-3mm}
			\includegraphics[width=\textwidth]{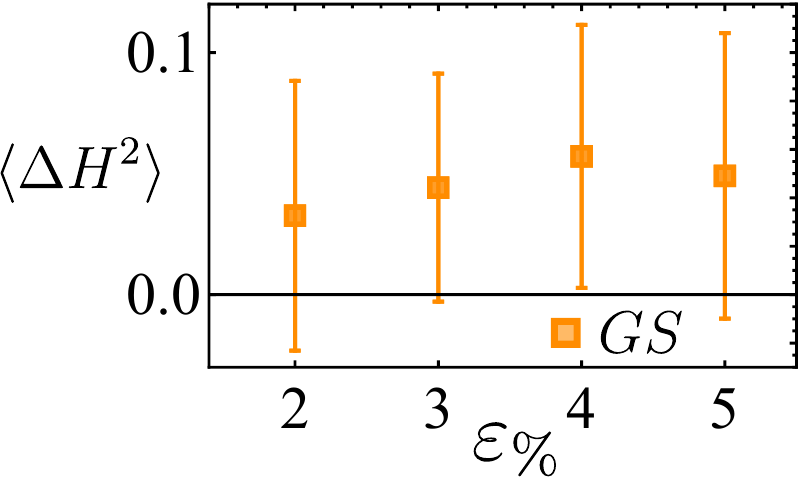}
			\label{fig:H2_vs_eps_teor}
		\end{subfigure} 
		\captionsetup{justification=centerlast}
		\caption{
			\label{fig:prot_teor}
			(a) SSH model ground state for $L=8$ and $\ell=L/2$, in the trivial (left) and in the topological (right) phases. Thicker lines indicate the larger interaction terms and thus where dimers form. Thus, the purity is $\simeq 1/2\,(1)$ in the topological (trivial) phase.
			(b) Purity for different values of $L$ obtained with the {rotation protocol $\mathbf{R}^*$ ignoring the interactions and every error source.} We consider the topological ground state for $L=8,\,12$ (squares) and the trivial one for $L=6,\,10$ (triangles). The dashed lines represent the expected values corresponding to $\mathrm{Tr}[\rho_\ell^2]$ for $\ell=L/2,\,L/2+1$. The error bars are the standard deviation computed over 20 repetitions of the protocol. 
			(c) Estimation of the energy variance renormalized with respect to the ground state energies for each value of $L$. 
			(d)-(e) Estimation of the purity and the Hamiltonian variance for $L=8$ by applying {the rotation protocol $\mathbf{R}^*$ with random fluctuation for different  relative amplitude of the fluctuations $\varepsilon_\%$.} The fluctuations change at each unitary transformation sample. We set  $N_U = 100$.
        }
	\end{figure}

	The model ground state exhibits two phases, a topological one for $|J_e|\gg |J_o|$, with localized edge excitations, and a trivial one for $|J_o|\gg |J_e|$. In both phases, the ground state bulk is composed by separable nearest-neighbors dimers sharing one excitation. The dimers form on those sites connected by the stronger interaction term leading, in the topological phase, to the localized boundary excitations. Figure \ref{fig:sshgs} shows the two phases for the SSH model ground state, with thicker lines indicating the stronger coupling.
Given a system bipartition with size $\ell$, the purity of the reduced density matrix can assume two values: they are $p_s=1$, if the boundary does not cross any dimer and $\rho_\ell$ describes a pure state, and  $p_d=1/2$, if the boundary crosses a dimer: measuring the purity of a given subsystem allows to distinguish the topological from the trivial phase.
	Moreover, RMs can also be used to extract the quantized topological invariants~\cite{Pollmann2012,Elben2020_SPT}.
	
	Finally, we  benchmark the ground state preparation by measuring the Hamiltonian variance
	\begin{equation}\label{eq:fluct}
	\langle \Delta  H_{mod}^2\rangle
	 =\mathrm{Tr}[\rho  H_{mod}^2]
	 -\mathrm{Tr}[\rho  H_{mod}]^2.
	\end{equation}
	In the results, we show the renormalized quantity $\langle\Delta  H^2\rangle=\langle\Delta  H_{mod}^2\rangle/\langle H_{mod}^2\rangle$. We first consider the exact ground state $|GS\rangle$, for which ideally we would measure $\Delta  H_{mod}^2=0$,  neglecting the effects due to the state preparation process. Then, we benchmark the rotation protocol on the state $|\widetilde{GS}\rangle= {P}[\ket{\downarrow \dots \downarrow}]$, where ${P}$ is the adiabatic state preparation protocol presented in \cite{DeLeseleuc2019} and the initial state is a fully ferromagnetic one.
    We also compare the variances obtained for the ground states $|GS\rangle$ and $|\widetilde{GS}\rangle$	with that of a separable anti-ferromagnetic state $|AF\rangle=\ket{\uparrow\downarrow\dots}$. 

	\begin{figure}
		\begin{subfigure}{0.95\columnwidth}
			\captionsetup{position=top,singlelinecheck=off,justification=raggedright}
			\caption{}
			\centering
			{\vskip -6mm}
			\includegraphics[width=\textwidth]{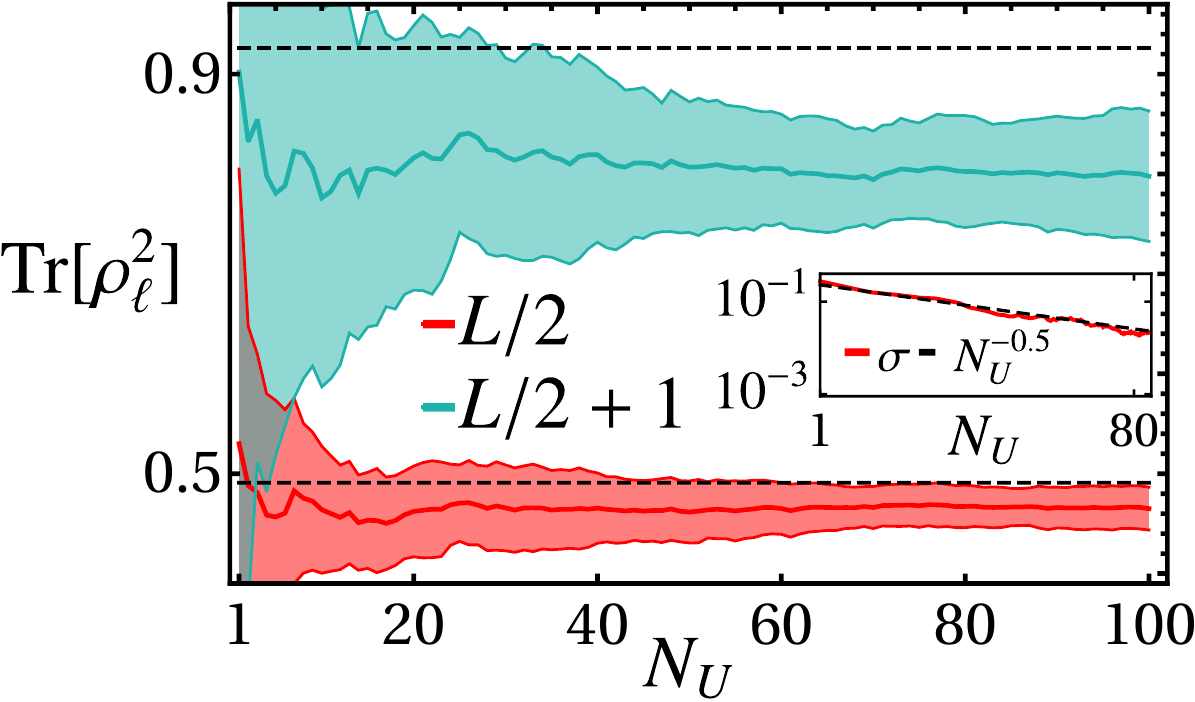}
			\label{fig:tr2_8}
		\end{subfigure} 
		{\vskip -2mm}
		\begin{subfigure}{0.95\columnwidth}
			\captionsetup{position=top,singlelinecheck=off,justification=raggedright}
			\caption{}
			\centering
			{\vskip -6mm}
			\includegraphics[width=\textwidth]{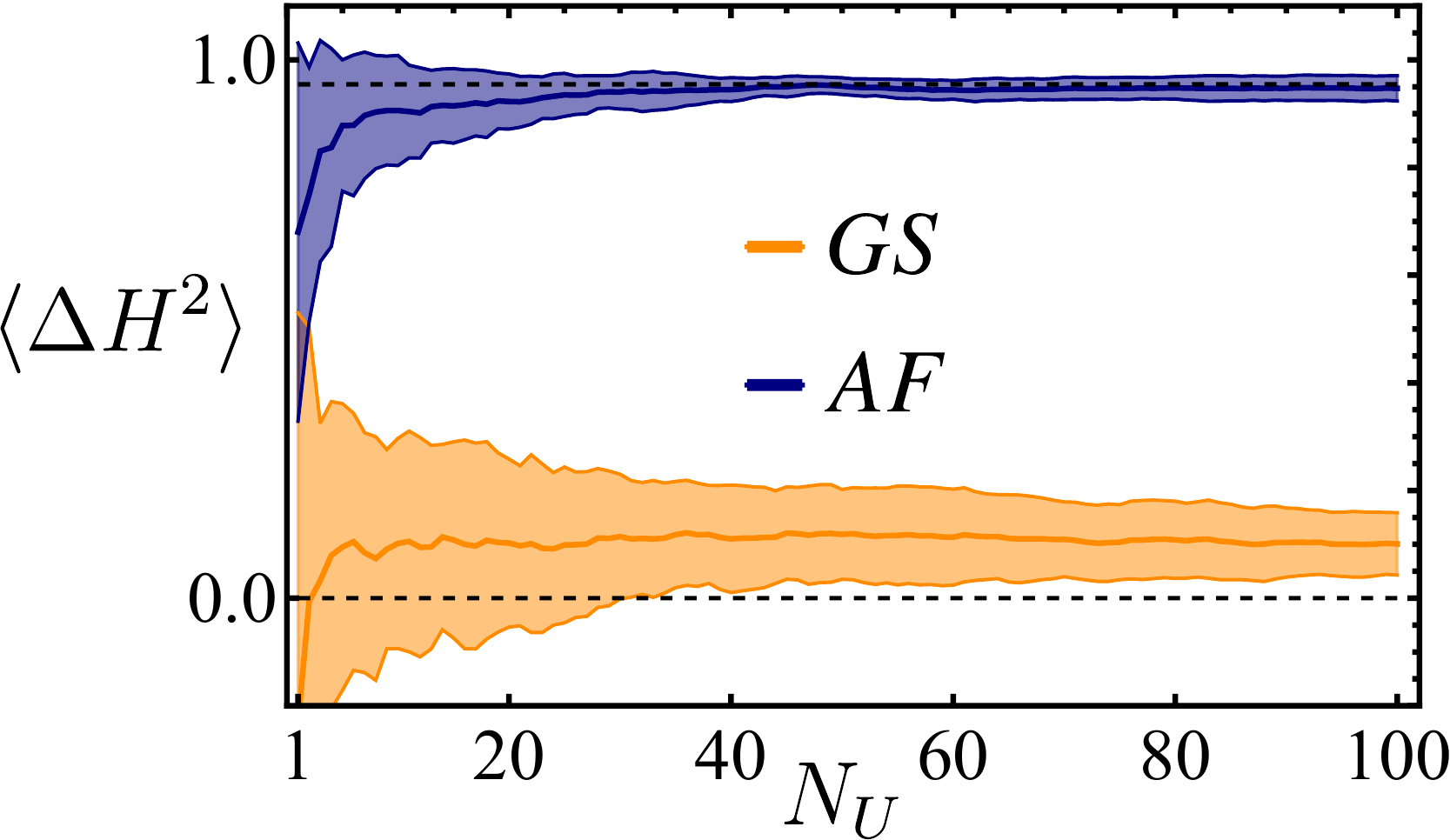}
			\label{fig:H2_8}
		\end{subfigure} 
		\captionsetup{justification=centerlast}
		\caption{
			\label{fig:L_8_full}
			The experimental protocol is applied to the SSH ground state $|GS\rangle$ to estimate the purity (a)  and  the Hamiltonian variance (b) as a function of the number of sampled unitaries $N_U$.
			We consider the topological phase and $L=8$. We plot the average  over 20 repetitions of the full estimation process (for each repetition we prepare the target state and evolve under ${H}_{tot}(t)$ for $N_{meas}\times N_U$ times). The colored areas correspond to the standard deviations.
			The inset shows square-root decreasing of the standard deviation of $\mathrm{Tr}[\rho_\ell^2]$ for $\ell=L/2$. Analogous behaviors are observed for $\ell=L/2+1$ and for the energy fluctuations. 
			The Hamiltonian variances are computed both for the ground state and for a separable, antiferromagnetic state. All energy variances are renormalized with respect to the respective ground state energies. The dashed lines show the exact values for the purities (a) and the variances (b). To estimate the probabilities we set $N_{meas}=400$. We set the relative variance of pulse fluctuations $\varepsilon_\%=3\%$.
		}
	\end{figure}


\subsection{Numerical results}
	We benchmark now the set of rotations $\mathbf{R^*}$. We start our analysis by considering the exact ground-state, with three different scenarios. In the first, we apply the rotation protocol by evolving the ground-state with the pulses shown in Fig. \ref{fig:totprot}. We consider only the Hamiltonian ${H}_{prot}$, thus for the moment ignoring interactions.
	Moreover, we do not add any fluctuations and the probabilities $P_U(s)$ are computed exactly. We compute the exact ground state $|GS\rangle$ for different sizes $L$ and estimate the purity $p_2$ of the reduced density matrix $\rho_\ell$ for a bipartition with sizes $\ell=L/2,L/2+1$ (see Fig. \ref{fig:tr2_vs_L_teor}).
	We fix the number of global unitary samplings $N_U=100$ and repeat the whole process $N_{ave}=20$ times. All the results shown hereafter are averaged over these repetitions and the error bars are estimated by taking the standard deviation. 
	In Fig. \ref{fig:tr2_vs_L_teor} the two expected values of the purity $p_d$ and $p_s$ are distinguishable within the error bars for all the values of $L$. 
	The colors indicate the size $\ell$ of the bipartition, while we use squares (triangles) for $L=8,12(6,10)$ to indicate that the ground state has been prepared in the topological (trivial) phase.
	The error bars we observe show the robustness of our protocol for different system size for the chosen number $N_U$. 
	Furthermore, we estimate energy fluctations, as shown in \ref{fig:H2_vs_L_teor}.  Note that the relations of Eqs.~\eqref{eq:Tr2} and \eqref{eq:rhoO} do not satisfy physical bounds, such as $1\geq\mathrm{Tr}[\rho^2]>0$ or positivity, therefore purity values larger than one or negative fluctuations can be encountered.
	
	We then add Gaussian pulse fluctuations with different amplitudes to check the impact of noise in the protocol implementation. We fix $L=8$ and repeat the estimations shown above. We still do not consider interactions and compute the probabilities $P_U(\mathbf{s})$ exactly. 
	The results shown in Fig.~\ref{fig:tr2_vs_eps_teor} for the purity and in Fig.~\ref{fig:H2_vs_eps_teor} for the Hamiltonian variance respectively are in good agreement with the expected values. The errorbars concerning the purity estimations allow to distinguish between $p_d$ and $p_s$.
These simulations provide a fundamental benchmark for future experiments, as they prove the robustness of the toolbox against imperfect pulse realizations.

	Finally, we simulate the  experimental protocol, evolving the ground state with the full Hamiltonian ${H}(t)$ defined in Eq.~\eqref{eq:tot_ham}, thus, considering the Rydberg interactions. The Hamiltonian ${H}_{prot}(t)$ is perturbed with Gaussian pulse fluctuations at each repetition of the protocol. Moreover, we estimate the probabilities $\tilde{P}_U(\mathbf{s})$  by simulating the measurement process and including measurement errors.
	In Figure \ref{fig:L_8_full}, we fix $L=8$ and estimate (a) the purity and (b) Hamiltonian variance  for a maximum number of unitary transformation samples $N_U=100$. For each unitary sample, we repeat the process $N_{meas}=400$ times to estimate the state probability amplitudes $\tilde{P}_U(\mathbf{s})$ in the measurement basis. {The values we obtain for the purity are well separated within the error bars. Thus, they allow us to distinguish between partitions with odd and even size, at a fixed state, or alternatively between the topological and trivial phase at fixed partition size}.
	Note that the effect of noise during the measurements tends to slightly reduce the estimation of the purity, i.e. the state appears as more mixed compared to a perfect measurement sequence, as a consequence of decoherence. If needed, this effect can be removed using rescaled probabilities based on calibration  experiments~\cite{Vermersch2018,Vovrosh2021,Satzinger2021}.
	The inset in Fig.~\ref{fig:tr2_8} shows the expected inverse square-root scaling of the standard deviation of the estimated purities. 
	Note that, when measuring purities, we take into consideration the statistical bias occurring from the estimation procedure of the probabilities $\tilde{P}_U(\mathbf{s})\tilde{P}_U(\mathbf{s'})$~\cite{Elben2019}.  
 {In particular, the unbiased purity $\mathrm{Tr}[\rho^2] = x\, N_{meas}/(N_{meas}-1) - 2^\ell/(N_{meas}-1)$, where $x$ is the biased result obtained from directly inserting the estimate $\tilde{P}$ into Eq.~\eqref{eq:Tr2}}.
    	\begin{figure}
		\begin{subfigure}{0.49\columnwidth}
			\captionsetup{position=top,singlelinecheck=off,justification=raggedright}
			\caption{}
			\centering
				{\vskip -4mm}
			\hspace{-3mm}
			\includegraphics[width=\textwidth]{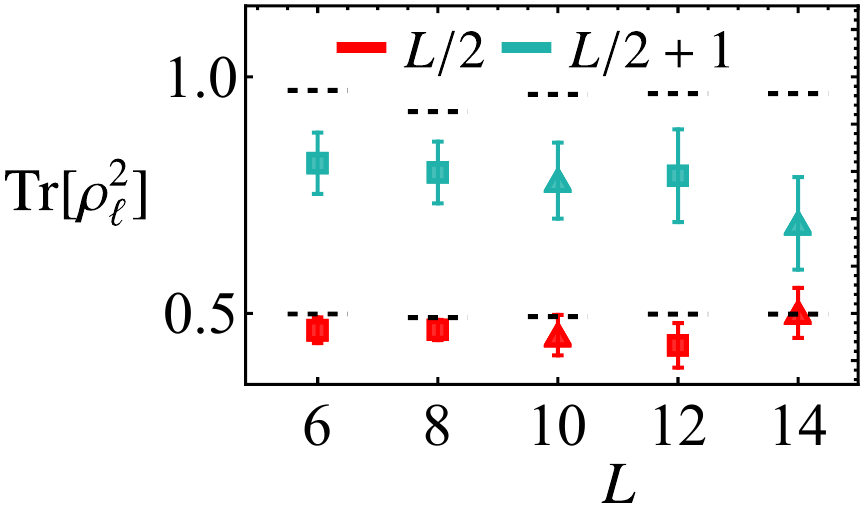}
			\label{fig:tr2_6_8}
		\end{subfigure} 
		\begin{subfigure}{0.49\columnwidth}
			\captionsetup{position=top,singlelinecheck=off,justification=raggedright}
			\caption{}
			\centering
				{\vskip -4mm}
			\hspace{-3mm}
			\includegraphics[width=\textwidth]{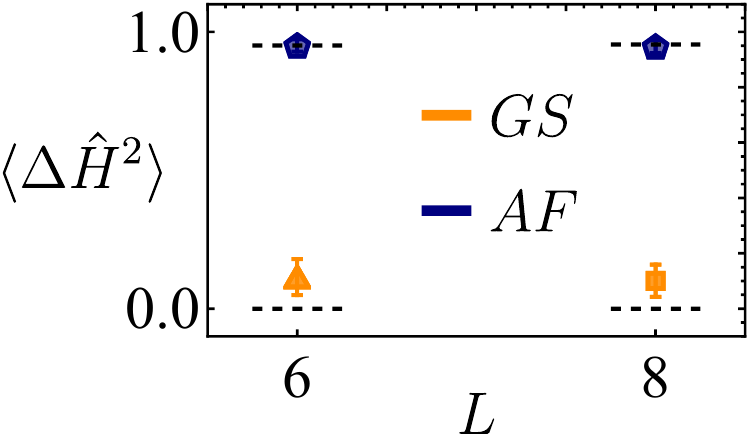}
			\label{fig:H2_6_8}
		\end{subfigure} 
		\begin{subfigure}{0.49\columnwidth}
			\captionsetup{position=top,singlelinecheck=off,justification=raggedright}
			\caption{}
			\centering
				{\vskip -4mm}
			\hspace{-3mm}
			\includegraphics[width=\textwidth]{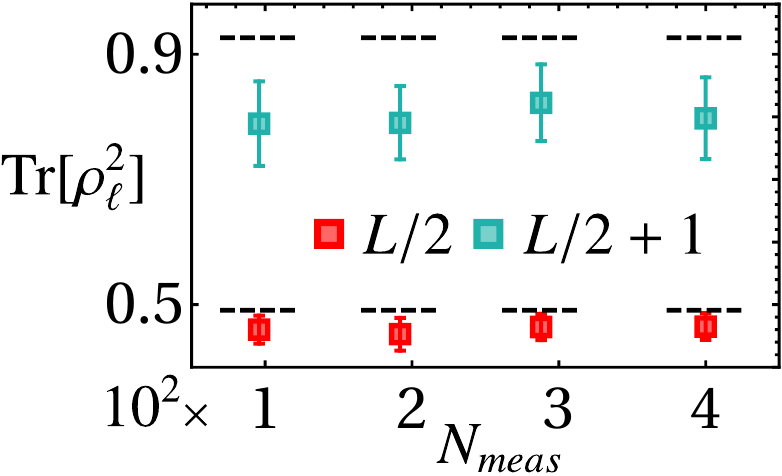}
			\label{fig:tr2_8_Nmeas}
		\end{subfigure} 
		\begin{subfigure}{0.49\columnwidth}
			\captionsetup{position=top,singlelinecheck=off,justification=raggedright}
			\caption{}
			\centering
				{\vskip -4mm}
			\hspace{-3mm}
			\includegraphics[width=\textwidth]{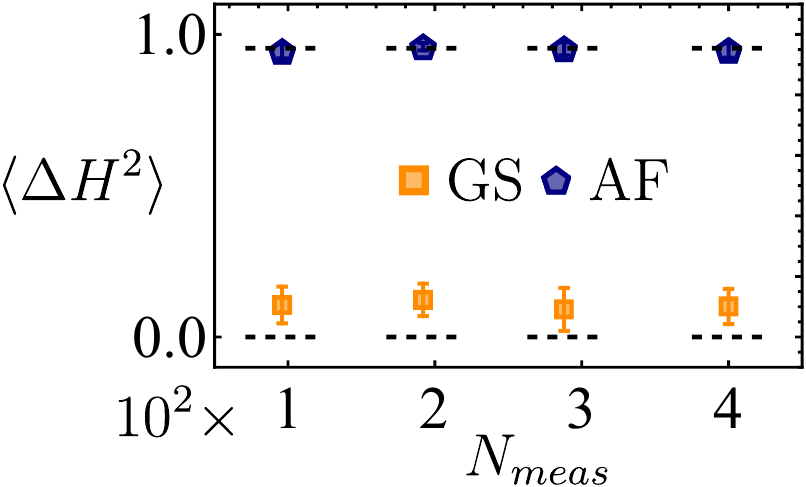}
			\label{fig:H2_8_Nmeas}
		\end{subfigure} 
		
		\captionsetup{justification=centerlast}
		\caption{
			\label{fig:comp6_8_100_400}
			(a) Estimation of the purity {(up to $L=14$)} and (b) the energy variance ($L=6,8$) via the experimental protocol. The systems is respectively prepared in the trivial phase ($L=6, 10, 14$)  phase) and in the topological one ($L=8,12$). Squares and triangles are used to distinguish between the topological and trivial SSH ground states while pentagons refer to the antiferromagnetic state.
			(c)-(d) Test of the experimental protocol robustness as a function of $N_{meas}$ for $L=8$. $L=6,8:N_U=100,N_{meas}=400$. $L=10,12,14:N_U=50, N_{meas}=800.$ $\varepsilon_\%=3 \%$.
		}
	\end{figure}
    
	In the estimation of the Hamiltonian variance, we compare the value relative to the model ground state with that of an antiferromagnetic separable state.  The obtained values are separated within the error bars, confirming once again the robustness of our RM toolbox to distinguish between states with different properties. We execute this procedure for $L=6, \dots, 14$, as shown in Figs.~\ref{fig:tr2_6_8} and \ref{fig:H2_6_8}. 
	Furthermore, we benchmark our toolbox by changing the number of measurements $N_{meas}$ for each global unitary sample. We consider the cases with $L=8$. In Figs.~\ref{fig:tr2_8_Nmeas} and \ref{fig:H2_8_Nmeas}, we observe that the results obtained for different values of $N_{meas}$ do not differ substantially, allowing us to consistently reduce the number of required iterations of the protocol for small lattice sizes.
	
\begin{figure}
		\begin{tabular}{cc}
		\begin{subfigure}{0.49\columnwidth}
			\captionsetup{position=top,singlelinecheck=off,justification=raggedright}
			\caption{}
			\centering
			{\vskip -4mm}
			\hspace{-3mm}
			\includegraphics[width=\columnwidth]{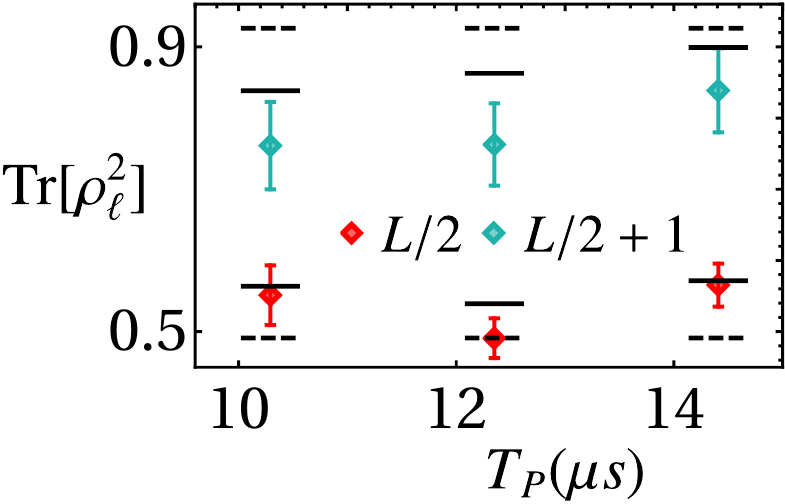}
			\label{fig:tr2_8_GS_exc}
		\end{subfigure} 
		     &  
		    \begin{subfigure}{0.49\columnwidth}
			\captionsetup{position=top,singlelinecheck=off,justification=raggedright}
			\caption{}
			\centering
			{\vskip -4mm}
			\hspace{-3mm}
			\includegraphics[width=\columnwidth]{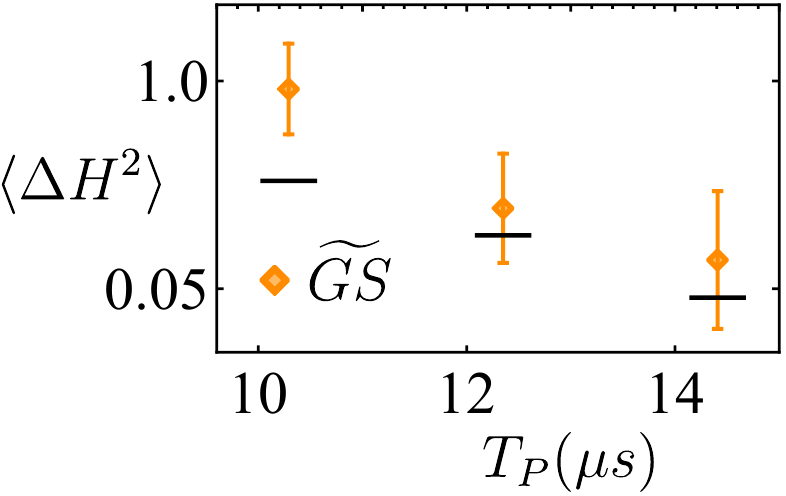}
			\label{fig:H2_8_GS_exc}
		\end{subfigure} 
		\end{tabular}
		\captionsetup{justification=centerlast}
		\caption{
			\label{fig:L_8_GS_exc}
			The purity (a) and the Hamiltonian variances (b) are estimated for the state $|\widetilde{GS}\rangle= {P}[\otimes|\downarrow\rangle_i$, where ${P}$ is the protocol presented in \cite{DeLeseleuc2019}, as a function of the preparation time $T_{P}$ for $L=8,\,N_{U}=100,\,N_{meas}=400,\,\varepsilon_\%=3$. In (a), continuous lines show the expected purities relative to  $|\widetilde{GS}\rangle$, while the dashed ones correspond to the values relative to the exact ground state. In the log-scale plot (b), the continuous lines show the expected variances relative to  $|\widetilde{GS}\rangle$.
		}
	\end{figure}

	{As an application of the measurement of the energy variance, we show that the protocol can be used to check the adiabatic preparation of the ground state.} We consider the adiabatic sequence presented in \cite{DeLeseleuc2019}, suitably scaled to the interaction strength adopted here. We obtain an imperfect ground state $|\widetilde{GS}\rangle$ for different preparation times $T_P$ and compute the purities and the Hamiltonian variances for $L=8$. In Figure~\ref{fig:L_8_GS_exc}, we report the purities (\ref{fig:tr2_8_GS_exc}) and the Hamiltonian variances (\ref{fig:H2_8_GS_exc}) relative to the experimental numerical simulation as a function of $T_P$. In Fig.~\ref{fig:tr2_8_GS_exc}, the dashed lines represent the values for the purities computed on the exact ground state $|GS\rangle$, while the continuous ones correspond to the imperfect ground state $|\widetilde{GS}\rangle$. As expected, the preparation protocol is not adiabatic for shorter processes, and the ground state properties are affected. The low interaction strengths we considered here require $\sim 10 \mu s$ to prepare the ground state. On the one hand, such a time scale also requires to consider incoherent effects. On the other hand, optimal control techniques allow to go beyond adiabatic protocols for state preparations and adopt faster ones~\cite{PhysRevA.84.012312,Caneva_2012}.
 
\begin{figure}
			\centering
			\includegraphics[width=0.9\columnwidth]{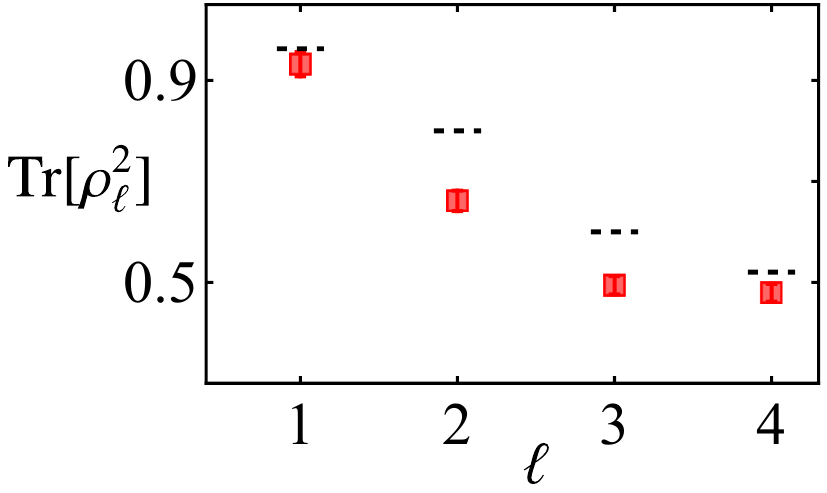}
		\caption{
			\label{fig:tr2_vs_ell_8}
			The purity respect to the bipartition of size $\ell$ is estimated for a state evolved for $T=1\mu s$ after a sudden quench. The dashed lines show the exact values. Numerical parameters: $L=8$, $N_U = 50$, $N_{meas} = 800$, $N_{ave}=10$.
		}
	\end{figure}

     {Finally, we test the our protocol in a different scenario.  First, we set $J_e = -J_o = 0.18\mathrm{MHz}$ in the Hamiltonian in Eq.(\ref{eq:tot_ham}), implementing a staggered XY model. Then, we consider the separable $L=8$ state with all spins down, except a single spin up the middle of the lattice. By evolving the system for a time $T=1\mu\mathrm{s}$, the domain wall spreads over the lattice until it reaches the boundary of the lattice. We measure the purity for different bipartitions of the lattice. The purity values estimated for different subsystem sizes $\ell = 1, ..., L/2$ are shown in Figure \ref{fig:tr2_vs_ell_8}. The values obtained from the protocol are compared with the exact one, represented by the dashed lines. We notice that the estimated values capture the entanglement growth as the subsystem boundary is shifted toward the center of the lattice. This result shows that our protocol can be used to infer different entanglement profiles in lattice models.}

	\section{Conclusion}
	
	{We have proposed a protocol to implement simultaneous local, independent unitary rotations in an interacting Rydberg quantum simulator. To benchmark it, we have investigated the ground state properties of the SSH chain and the XY model after a sudden quench. 
 We have shown the effectiveness of the local random unitary rotations protocol, despite the presence of interactions. 
 It allows us to estimate quantities such as the purity of a system bipartition and the Hamiltonian variance, taking into account realistic experimental parameters, the influence of residual interactions, and imperfections such as finite read-out fidelities.}
	The presented results provide a complete RM toolbox to probe entanglement \cite{Elben2018,Brydges2019,Elben2020_Mixed}, many-body topological invariants \cite{Elben2020_SPT} and quantum state fidelities \cite{Elben2020_xPlatform}, but also to measure any quantum observable from classical shadows \cite{Huang2020},  in Rydberg quantum simulators and quantum computers.
	\section{Acknowledgements}
	
We thank Peter Zoller for inspiring discussions and comments on the manuscript. 
BV acknowledges funding from the Austrian Science Fundation (FWF, P 32597 N), and from the French National Research Agency (ANR-20-CE47-0005, JCJC project QRand).
	AE acknowledges funding by the German National Academy of Sciences Leopoldina under the grant number LPDS 2021-02 and by the Walter Burke Institute for Theoretical Physics at Caltech.
	SN and SM acknowledge support from the EU Horizon 2020 research and innovation program (PASQuanS2 and EuRyQa), the QuantERA projects QuantHEP and T-NISQ, the NextGenerationEU project CN00000013 - Italian Research Center on HPC, Big Data and Quantum Computing, and the WCRI-Quantum Computing and Simulation Center of Padova University.
TL and AB acknowledge support from the EU Horizon 2020 research and innovation program 
under grant agreement no. 817482 (PASQuanS), the Agence Nationale de la Recherche (ANR, project RYBOTIN).
SN acknowledges that this project has received funding from the European Union’s Horizon Europe research and innovation
program under the Marie Skłodowska-Curie grant agreement No. 101059826 (ETNA4Ryd). SN acknowledges support from the Harvard Department of Physics.
	
\bibliography{RM_bibliography}

\end{document}